\documentclass{ws-mpla}

\usepackage{amsmath,amssymb,amsbsy}
\usepackage{graphicx}
\usepackage{epsfig}

\newcommand*{\chpt}{\raise0.4ex\hbox{$\chi$}PT}
\newcommand*{\schpt}{S\raise0.4ex\hbox{$\chi$}PT}
\newcommand*{\ie}{\textit{i.e.},\ }

\newcommand*{\etc}{\textit{etc.}}

\newcommand*{\MeV}{{\rm Me\!V}}

\newcommand*{\Tr}{\ensuremath{\operatorname{Tr}}}
\newcommand*{\tr}{\ensuremath{\operatorname{tr}}}

\newcommand{\semitimes}{\mathrel>\joinrel\mathrel\triangleleft}
\def\gtwid{{\,\raise.35ex\hbox{$>$\kern-.75em\lower1ex\hbox{$\sim$}}\,}}
\def\ltwid{{\,\raise.35ex\hbox{$<$\kern-.75em\lower1ex\hbox{$\sim$}}\,}}
\def\leftvec{{\raise1.5ex\hbox{$\leftarrow$}\kern-1.00em}}
\def\rightvec{{\raise1.5ex\hbox{$\rightarrow$}\kern-1.00em}}
\def\half{{\scriptstyle \raise.2ex\hbox{${1\over2}$}}}
\def\threehalves{{\scriptstyle \raise.15ex\hbox{${3\over2}$}}}
\def\third{{\scriptstyle \raise.15ex\hbox{${1\over3}$}}}
\def\third{{\scriptstyle \raise.15ex\hbox{${1\over3}$}}}
\def\twothirds{{\scriptstyle \raise.15ex\hbox{${2\over3}$}}}
\def\fourth{{\scriptstyle \raise.15ex\hbox{${1\over4}$}}}

\newcommand{\Dslash}{\ensuremath{D\!\!\!\! /\; }}

\newcommand{\cD}{\ensuremath{\mathcal{D}}}

\newcommand{\cL}{\ensuremath{\mathcal{L}}}
\newcommand{\cM}{\ensuremath{\mathcal{M}}}

\newcommand{\cO}{\ensuremath{\mathcal{O}}}
\newcommand{\cP}{\ensuremath{\mathcal{P}}}

\newcommand{\cU}{\ensuremath{\mathcal{U}}}
\newcommand{\cV}{\ensuremath{\mathcal{V}}}

\newcommand{\eq}[1]{Eq.~\eqref{eq:#1}}

\newcommand{\eqs}[2]{Eqs.~\eqref{eq:#1} and \eqref{eq:#2}}

\begin{document}

\markboth{C.~Aubin}
{Current Physics Results from Staggered Chiral Perturbation Theory}

\catchline{}{}{}{}{}

\title{Current Physics Results from Staggered Chiral Perturbation Theory}

\author{\footnotesize C.~AUBIN}

\address{Physics Department,
        Columbia University \\
        New York, New York 10027, USA\\
        caubin@phys.columbia.edu}

\maketitle

\pub{Received (Day Month Year)}{Revised (Day Month Year)}

\begin{abstract}
We review several results that have been obtained using lattice QCD with the staggered quark formulation. Our focus is on the quantities that have been calculated numerically with low statistical errors and have been extrapolated to the physical quark mass limit and continuum limit using staggered chiral perturbation theory. We limit our discussion to a brief introduction to staggered quarks, and applications of staggered chiral perturbation theory to the pion mass, decay constant, and heavy-light meson decay constants.

\end{abstract}

\ccode{PACS Nos.: 11.15.Ha,11.30.Rd,12.38.Gc,12.39.Fe}

\section{Introduction}

Currently, a large number of lattice results have appeared using staggered fermions to discretize the quark fields.\cite{Davies:2003ik,Aubin:2004fs,Allison:2004be,Aubin:2004ej,Aubin:2005ar,MILCSugar} What contrasts these calculations with earlier lattice calculations is the use of 2+1 dynamical flavors of light quarks, with up and down quark masses down to $\sim m_s/10$ (where the lattice strange quark mass is roughly its correct value), multiple lattice spacings, and the ability to understand the light quark and continuum limit. This last  step allows one to minimize systematic errors in many of these results, so as to make an accurate extrapolation to the physical values of the light quark masses. 

For staggered fermions, this step is highly non-trivial. Introductions of scaling violations which are rather large at finite lattice spacings cause strong effects. These violations are reduced using improved actions, but are still not negligible.\cite{Orginos:1998ue} Because many of these violations add to the squared pion mass, it is perhaps not surprising that when attempting to construct the relevant chiral perturbation theory, they must be included as chiral symmetry breaking terms, similar to the way the quark masses enter.

This review is intended to be focussed on this aspect of analyzing lattice data. We begin by outlining lattice QCD with staggered quarks in Sec.~\ref{sec:latQCD}. We continue by discussing the general procedures for deriving a low-energy effective theory for QCD with staggered quarks in Sec.~\ref{sec:schpt}. We then use this to calculate some pion-related quantities in Sec.~\ref{sec:pions}, and include heavy quarks in Sec.~\ref{sec:heavylights}. Overall, we will focus primarily on results from chiral perturbation theory that have been applied to lattice data to obtain relatively small systematic errors as a result. There are many other quantities that have been studied, and we will briefly touch upon them in Sec.~\ref{sec:conc}.

\section{Lattice QCD with Staggered Quarks}\label{sec:latQCD}

We begin by formulating Quantum Chromodynamics on a discrete spacetime lattice with a uniform lattice spacing $a$. The quark fields live on sites, and the gluon fields live on links, although we will not discuss the gluons in detail. While focussing on the quarks, we will discuss everything in terms of the free theory for simplicity, but the results carry over to the interacting case. 

The simplest discretization is given by replacing the derivatives in the free Euclidean quark action by difference operators:
\begin{equation}\label{eq:naive_action}
	S_\textrm{quarks} = a^4\sum_{f,x}\left[
	\frac{1}{2a}
	\sum_\mu \gamma_\mu \bar Q_{f,x} 
	\left( Q_{f,x + a\hat\mu} - Q_{f,x -a\hat\mu}\right)
	+ m_f\bar Q_{f,x} Q_{f,x}
	\right]\ ,
\end{equation}
where $f$ runs over the three lightest quarks $u,d,$ and $s$,\footnote{The heavy quarks play no dynamical role and can be integrated out at the scales usually used in lattice simulations.} $x$ is a spacetime point, $\hat\mu$ is a unit vector in the $\mu$-direction, and $m_f$ is the mass of the quark flavor $f$. This action has the well-known doubling problem, which can be seen in the momentum-space propagator for a given flavor $f$
\begin{equation}\label{eq:naive_prop}
	a\frac{-i\sum_\mu \gamma_\mu \sin(a p_\mu) + am_f}
	{\sum_\mu \sin^2(a p_\mu) + (am_f)^2}\ .
\end{equation}
This has a low-energy mode as $a\to 0$ at $p=0$, but also when any component of $p$ is near $\pi/a$ or 0---at one of the corners of the Brillouin Zone. The appearance of these additional 15 doublers is a general result of discretizing fermions.\cite{Nielsen:1981xu,Nielsen:1980rz} There several techniques to rid oneself of the doubling problem,\cite{Wilson:1975hf,Susskind:1976jm,Ginsparg:1981bj,Kaplan:1992bt,Shamir:1993zy} and we will focus here on the staggered discretization.\cite{Susskind:1976jm}

To arrive at the staggered quark formalism, we realize that when going to a discretized Euclidean space, one breaks the $SO(4)$ rotational symmetry down to a discrete subgroup composed of finite rotations by $\pi/2$ (denoted by $SW_4$). This implies that we can perform a redefinition of the quark fields in \eq{naive_action} that diagonalizes the $\gamma$ matrices, thereby decoupling the different spinor components of $Q_f$. We then keep only one spinor component per field and we have decreased the number of degrees of freedom by a factor of four, and giving us the action
\begin{equation}\label{eq:stag_action}
	\cL_\textrm{quarks} = a^4\sum_{f,x}\left[
	\frac{1}{2a}
	\sum_\mu \eta_{\mu,x} \bar \psi_{f,x} 
	\left( \psi_{f,x + a\hat\mu} - \psi_{f,x -a\hat\mu}\right)
	+ m_f\bar \psi_{f,x} \psi_{f,x}
	\right]\ ,
\end{equation}
where $\eta_{\mu,x} = \sum_{\nu<\mu}(-1)^{x_\nu}$, and $\psi_f$ is the single-component field remaining after the field redefinition and dropping the other spinor components.
To show that this set of 16 single-component fermions (per flavor $f$) corresponds to four four-component Dirac fermions is a straightforward exercise that we will not pursue here. The most important point here is that different species (or ``tastes'') of a given flavor are identified (in momentum space) with different corners of the Brillouin Zone.\cite{Golterman:1984cy}

It is possible to take only a single flavor and add a more general mass matrix which can split the degeneracy in \eq{stag_action} so as to describe four-flavor QCD (the four tastes are identified as four different flavors).\cite{Golterman:1984cy} However, this procedure has technical difficulties making it rather tough to implement numerical. What is done in practice is to create one staggered field for each flavor one wishes to simulate. After evaluating the path integral over the fermions, we obtain the fermion determinant for each staggered field, which is a determinant describing four tastes. Since these four tastes are degenerate in mass, the fourth-root of this determinant should then be a determinant describing a single taste of the staggered quark.\cite{Marinari:1981qf}

\begin{figure}[htbp]
\begin{center}
\includegraphics[width=4in]{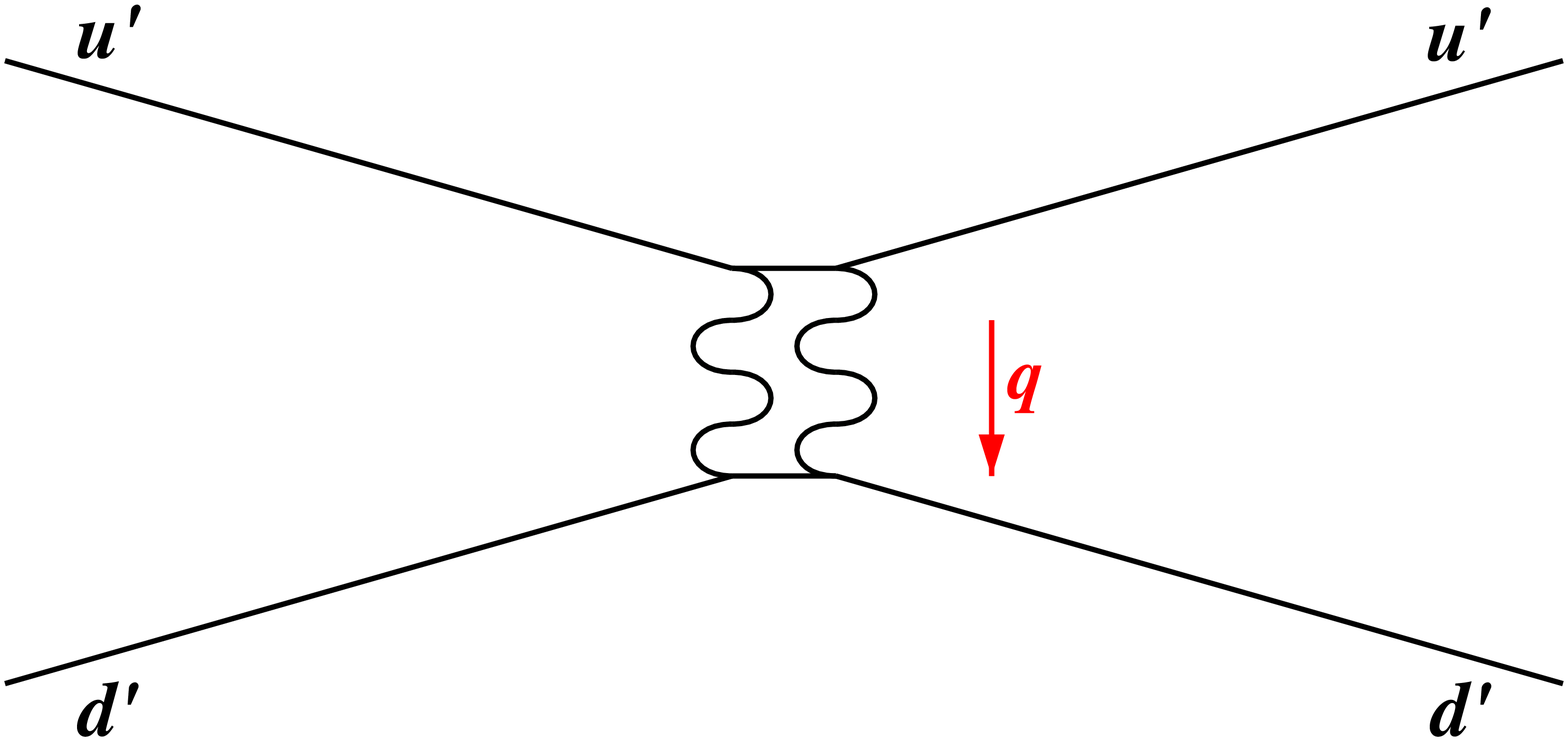}
\caption{A four-quark diagram which can introduce taste-violating interactions due to the exchange of high-momentum gluons. $q$ here
is assumed to be of order $\pi/a$. $u$ and $u'$ are up quarks of different tastes, as are $d$ and $d'$.}
\label{fig:fourquark}
\end{center}
\end{figure}

This ``rooting'' technique is rather common in staggered simulations, but it is not without theoretical issues. At finite lattice spacing, there are interactions among staggered quarks which violate the taste symmetry: the symmetry which rotates the tastes among themselves and is a symmetry of the free theory. In the interacting theory, the exchange of high momentum gluons can change the taste of a quark. This is shown in Fig.~\ref{fig:fourquark} and one can see that by exchanging a momentum of order of the cutoff ($q\approx \pi/a$ in the figure) does not move the quark far off shell, as it would with other discretizations, but instead changes the quark to a different taste with a small amount of momentum. 

The question remains as to whether or not it is valid to take the fourth root of the staggered determinant \emph{before} one takes the continuum limit. If these two operations did not commute, then a rooted-staggered theory would not be equivalent to a one-flavor continuum theory in the continuum limit. Although there is no proof that exists that proves the validity of this procedure, there are indications that the rooted theory is legitimate. There is a large body of supporting evidence in favor of the validity of the rooted theory, as well as some evidence implying this trick is not valid\cite{Bernard:2006zw,Creutz:2006ys,Bernard:2006vv,Bernard:2006ee,Shamir:2006nj} (we refer the reader to Ref.~\refcite{Sharpe:2006re} for an excellent review of the current status of the rooting procedure). Assuming the validity of this trick, we will now pursue the application of \chpt\ to the rooted staggered theory.

Although the rooting procedure gives the correct number of degrees of freedom in the sea sector, this does not mean that we cannot see the effects of these additional tastes. In fact, we will see that staggered chiral perturbation theory, even after correcting for the fourth-root trick, shows a significant dependence on the different tastes of quarks that arise. 

\section{Staggered Chiral Perturbation Theory}\label{sec:schpt}

Since one cannot simulate at physical $u$ and $d$ quark masses currently, one must extrapolate to the physical masses from the larger masses actually used. For this one needs chiral perturbation theory (\chpt). However, continuum \chpt\ is not appropriate for a lattice theory, as there are other sources of chiral symmetry breaking besides the mass term. For staggered quarks, this additional chiral symmetry breaking arises primarily from the taste-violating interactions discussed in the previous section\footnote{There could be other forms of discretization errors but taste violations are dominant.}. As such, one must appropriately modify the chiral theory to take into account these additional chiral symmetry breaking effects,\cite{Lee:1999zx,Bernard:2001yj,Aubin:2003mg} and construct a staggered chiral perturbation theory (\schpt).

The first step in this procedure is to write down the Symanzik effective action\cite{Lee:1999zx} for the lattice theory. The idea is that, given a lattice action $S_\textrm{lat}$, one can write it as an expansion in $a$ about the continuum limit
\begin{equation}
	S_\textrm{lat} = \int d^4x\left[\cL_0 
	+ a \cL_1 + a^2 \cL_2 + \cdots\right]\ ,
\end{equation}
where $\cL_0$ is the continuum QCD Lagrangian. For the staggered quark and gluon actions, this expansion is in $a^2$ (there are no errors linear in $a$), so only even values of $n$ are included in this sum. We will work only to $\cO(a^2)$ in this review. 

The first term, $\cL_0$, is straightforward, as it is merely the continuum Lagrangian corresponding to the lattice theory. For higher-order terms, all possible operators that satisfy the lattice symmetries must be included, making this procedure progressively more difficult when working beyond $\cL_2$. We will discuss the $\cL_0$ and some of the $\cL_2$ terms in detail, although the $\cL_4$ terms have been worked out as well.\cite{Sharpe:2004is} In current lattice simulations with staggered quarks, we have $a^2\Lambda^2_{QCD}\sim m^2_\pi/\Lambda_{\chi}^2$ with $\Lambda_{\chi}$ the chiral scale. Since in the continuum chiral expansion, $p^2\sim m^2_\pi$, we use a power counting scheme such that $p^2\sim a^2$ as well. This implies that we will omit terms in $\cL_2$ that have powers of the quark masses or derivatives, as these terms would necessarily be of order $\cO(p^2 a^2)$ (we use $p$ generically for any power of derivative or pion mass for simplicity), which is one higher order than that which we are working to.

Once we have the Symanzik action worked out to a given order, we can then transcribe the Symanzik terms to the underlying chiral theory. This process is well understood for the continuum theory\cite{Gasser:1983yg,Gasser:1985gg} and we will work through the basic details when discussing $\cL_0$, as going through this process for the higher-order terms is a straightforward extension.

\subsection{$\cL_0$}

Although the $\cL_0$ term is simply the continuum Lagrangian, recall that this action is one with four tastes for every flavor of quark. For QCD with $N$ flavors, we have
\begin{equation}\label{eq:Lag0}
	\cL_0 = 
	\overline{q}_L(x) \Dslash q_L(x) + 
	\overline{q}_R(x) \Dslash q_R(x) + 
	\overline{q}_L(x) M q_R(x) + 
	\overline{q}_R(x) M^\dag q_L(x) + 
	\cdots\ ,
\end{equation}
where the $\cdots$ refers to terms which do not include the quark fields, such as the pure gluon Lagrangian. The quark fields are $4N$-component objects with the 4 tastes for each of the $N$ flavors, and $M$ is the $4N\times 4N$ mass matrix (note that the four tastes for each flavor have the same mass). Also, the mass matrix is real, but the forthcoming spurion analysis is simpler by considering it to be a complex operator; in the end we set $M=M^\dag$. $\cL_0$ has an $SU(4N)_L\times SU(4N)_R$ chiral symmetry in the limit $M\to0$, where, (with $L,R\in SU(4N)_{L,R}$)
\begin{equation}
	q_L \to L q_L ,\ q_R\to R q_R\ 
\end{equation}
is a symmetry of \eq{Lag0}.
The chiral symmetry is broken down to the vector subgroup $SU(4N)_V$ by a non-vanishing quark condensate, which gives rise to $(4N)^2-1$ Goldstone bosons. These bosons are encapsulated in a field $\Sigma = \exp\left[i\Phi/f\right]$, where 
\begin{equation}
	\Sigma \to L\Sigma R^\dag\ ,
\end{equation}
and $\Phi$ is the $4N\times 4N$ matrix which contains the Goldstone bosons. For the case where $N=3$, we have
\begin{eqnarray}\label{eq:Phi}
	\Phi = \left( \begin{matrix}
     		 U  & \pi^+ & K^+  \\*
     		 \pi^- & D & K^0  \\*
		 K^-  & \bar{K^0}  & S 
		 \end{matrix} \right),
\end{eqnarray}
where $U = \sum_{t=1}^{16} U_t T_t$, \etc, and $T_t\in\{\xi_5, i\xi_{\mu5}, i\xi_{\mu\nu}, \xi_{\mu}, \xi_I\}$. We use the Euclidean gamma matrices $\xi_{\mu}$, with
$\xi_{\mu\nu}\equiv \xi_{\mu}\xi_{\nu}$, $\xi_{\mu5}\equiv \xi_{\mu}\xi_5$, and $\xi_I \equiv I$ is the $4\times 4$ identity matrix. We write \eq{Phi} in the flavor basis, where $U \sim \bar u u,\ D\sim \bar d d,$ \etc 

Since the quark masses are non-zero but small compared with $\Lambda_{QCD}$, we can include them as a soft breaking of the chiral symmetry. To do this systematically, we promote the mass matrix to a spurion field with the transformation
\begin{equation}
	M\to L M R^\dag,\ M^\dag \to R M^\dag L^\dag \ ,
\end{equation}
which makes $\cL_0$ invariant under the chiral symmetry group. Thus, using $\Sigma, \Sigma^\dag, M, $ and $M^\dag$, we can write down a low-energy effective action for $\cL_0$ (after setting $M=M^\dag$):
\begin{equation}
	\cL^\chi_0 = 
	\frac{f^2}{8}\Tr\left(\partial_\mu \Sigma \partial_\mu 
	\Sigma^\dag\right) - \frac{\mu f^2}{4}\Tr\left(M\Sigma
	+\Sigma^\dag M\right) \ .
\end{equation} 
This is just a continuum \chpt\ expression for $4N$ flavors of quarks in Euclidean space. $f$ and $\mu$ are unknown parameters that are directly related to the tree-level pion decay constant and chiral condensate.

The methodology of determining the low-energy chiral Lagrangian from the Symanzik action is the same for $\cL_2$, $\cL_4$, and so forth. We need to know the operators that exist at the level of the Symanzik action, which can, in general, break the chiral symmetry of the massless QCD Lagrangian. By promoting objects that appear in the Symanzik action to spurion fields with specific transformation properties (as we did for the mass matrix), we can make the theory invariant and then from this, we can generate the corresponding operators at the chiral level.

\subsection{$\cL_2$}

Terms in the Symanzik action contributing to $\cL_2$ come from several sources. They can only arise from two-quark operators or four-quark operators, but we can immediately drop the two-quark operators from our analysis here. This is because a two-quark operator has dimension three, which goes to dimension one when we add in the required factor of $a^2$. This means we need three powers of mass, in the form of the quark mass or derivatives, and then the overall term will be at least of order $\cO(a^2p^4)$, which is higher order than that to which we are working. The four-quark operators are the only terms we need currently (one can easily see that higher numbers of quark fields will only contribute at higher powers of $a^2$). Effectively, we are looking at the terms in $\cL_2$ in the limit $p^2=0$ and $M=0$. 

In determining the operators that arise in $\cL_2$, one must first enumerate all of the operators that are consistent with the full lattice symmetry group at finite $a$. The full symmetry group was discussed for $N=1$ in Ref.~\refcite{Lee:1999zx} and for general $N$ in Ref.~\refcite{Aubin:2003mg}. For non-zero quark mass (assuming $n$ of the quarks are degenerate and $m$ are non-degenerate, with $N=m+n$), this symmetry group is $U(n)_\textrm{vec}\times\left(U(1)_\textrm{vec}\right)^m\times
 \Gamma_4\semitimes SW_{4,\textrm{diag}}$. Here, $U(m)_\textrm{vec}$ and $U(1)_\textrm{vec}$ are flavor number symmetries, $\Gamma_4$ is the Clifford group with  generators $\xi_\mu$, and $SW_{4,\textrm{diag}}\subset SO(4)$ is the group of hypercubic rotations in Euclidean space.

The number of operators that contribute to $\cL_2$ is around 25, where spin and taste indices are uncoupled (there are 10 operators where the spin and taste indices are coupled, but these do not contribute at this order). Many of these, however, do not need to be analyzed, as it was shown in Ref.~\refcite{Aubin:2003mg} that only ``odd-odd'' four-quark operators\footnote{Odd bilinears are those where the staggered quark fields are separated by an odd number of links on the lattice.} need to be included in our list. With this realization, we can take the analysis for the $N=1$ case from Ref.~\refcite{Lee:1999zx} and make the replacement of the taste matrices
\begin{equation}
	\xi_t \to \xi_t^{(N)} = \xi_t \otimes \mathbf{1}_N
\end{equation}
where $\mathbf{1}_N$ is the $N\times N$ identity matrix in flavor space, and $t$ can be any of the 16 tastes. 

Although we will not go into detail on the analysis for all of the operators, we will sketch the steps for a single set of operators. For simplicity we will pick a four-quark operator that has a vector spin structure and arbitrary taste structure:
\begin{equation}\label{eq:Operator}
	\cO_T = \sum_\mu\left[
	\overline{q}_R\left(\gamma_\mu\otimes T_R^{(N)}\right)q_R+
	\overline{q}_L\left(\gamma_\mu\otimes T_L^{(N)}\right)q_L
	\right]^2\ ,
\end{equation}
where the $q$'s and $\overline{q}$'s are the same $4N$-component objects above, and the taste matrices are going to be set equal in the end, $T_R^{(N)}=T^{(N)}_L=T^{(N)}$. $T^{(N)}$ can either be a pseudoscalar ($\xi_5^{(N)}$) or tensor ($\xi_{\mu\nu}^{(N)}$) taste, since this has to be composed of odd bilinears.\cite{Lee:1999zx} We give these taste matrices the spurion transformation under our $SU(4N)_L\times SU(4N)_R$ chiral symmetry:
\begin{equation}
	T_R \to RT_RR^\dag,\quad T_L \to LT_LL^\dag
\end{equation}
to make \eq{Operator} a chiral invariant, just as we did for the mass term in $\cL_0$. Since the only objects that appear in this term are the taste matrices and the quarks, we construct the chiral-level operator from $T_{R,L}$, $\Sigma$ and $\Sigma^\dag$. The only operator that arises is
\begin{equation}
	\cO_T^\chi \sim \Tr[T_L \Sigma T_R \Sigma^\dag]\ .
\end{equation}
This term leads to two different operators when we set $T_R^{(N)}=T^{(N)}_L=T^{(N)}$, with $T^{(N)}$ either the pseudoscalar or tensor tastes:
\begin{eqnarray*}
	\cO_{5}^\chi & \sim & \Tr[\xi^{(N)}_5 
	\Sigma \xi^{(N)}_5 \Sigma^\dag]\ , \\
	\cO_{\mu\nu}^\chi & \sim & \sum_{\mu<\nu}
	 \Tr[\xi^{(N)}_{\mu\nu} \Sigma \xi^{(N)}_{\nu\mu} \Sigma^\dag]\ .
\end{eqnarray*}

Working through the rest of the operators in Ref.~\refcite{Lee:1999zx}, we get a total of eight operators at the chiral level. We write these as
\begin{equation}
	a^2\cV_\chi = a^2\cU + a^2\cU^\prime
\end{equation}
with 
\begin{eqnarray}
	\label{eq:U}
	-\cU 
	& = & 
	  C_1 \Tr(\xi^{(N)}_5\Sigma\xi^{(N)}_5\Sigma^{\dagger})
    + C_3\frac{1}{2} \sum_{\nu}[ \Tr(\xi^{(N)}_{\nu}\Sigma
	\xi^{(N)}_{\nu}\Sigma) + h.c.] \nonumber \\*
	& & 
	+ C_4\frac{1}{2} \sum_{\nu}[ \Tr(\xi^{(N)}_{\nu 5}\Sigma
	\xi^{(N)}_{5\nu}\Sigma) + h.c.] 
	+ C_6\ \sum_{\mu<\nu} \Tr(\xi^{(N)}_{\mu\nu}\Sigma
	\xi^{(N)}_{\nu\mu}\Sigma^{\dagger}) \\*
	\label{eq:U_prime}
	-\cU^\prime 
	& = & C_{2V}\frac{1}{4} 
		\sum_{\nu}[ \Tr(\xi^{(N)}_{\nu}\Sigma)
	\Tr(\xi^{(N)}_{\nu}\Sigma)  + h.c.]
	+C_{2A}\frac{1}{4} \sum_{\nu}[ \Tr(\xi^{(N)}_{\nu
         5}\Sigma)\Tr(\xi^{(N)}_{5\nu}\Sigma)  + h.c.] \nonumber \\*
	& & +C_{5V}\frac{1}{2} \sum_{\nu}[ \Tr(\xi^{(N)}_{\nu}\Sigma)
	\Tr(\xi^{(N)}_{\nu}\Sigma^{\dagger})]
	 +C_{5A}\frac{1}{2} \sum_{\nu}[ \Tr(\xi^{(N)}_{\nu5}\Sigma)
	\Tr(\xi^{(N)}_{5\nu}\Sigma^{\dagger}) ]\ .
\end{eqnarray}

Thus, the full Lagrangian at the chiral level including both $\cL_0$ and $\cL_2$ in Euclidean space is
\begin{equation}\label{eq:ChiralAction}
	\cL^\chi = 
	\frac{f^2}{8}\Tr\left(\partial_\mu \Sigma \partial_\mu 
	\Sigma^\dag\right) - \frac{\mu f^2}{4}\Tr\left(M\Sigma
	+\Sigma^\dag M\right)
	+\frac{2m_0^2}{3}\Tr[\Phi]^2+ a^2\cV \ ,
\end{equation} 
where we have dropped terms that come from higher order in the joint $m^2,a^2$ expansion. We have also included here the term that is required by the anomaly (note that $\Tr[\Phi]$ is merely a sum of the flavor-neutral taste-singlet mesons), and we refer the reader to Ref.~\refcite{Gasser:1985gg} for more details. Gasser \& Leutwyler have worked out the terms that come in at $\cO(p^4,m^2p^2,m^4)$\cite{Gasser:1985gg} and Sharpe \& Van de Water have included the terms which arise at $\cO(p^2a^2, m^2a^2,a^4)$,\cite{Sharpe:2000bn} but we will not include these contributions to the action explicitly here. They will appear as analytic terms when we calculate quantities to one-loop order, however. 

\section{Pions}\label{sec:pions}

We can now use \eq{ChiralAction} to calculate light meson (referred to from now on generically as ``pions'') quantities. We first expand the field $\Sigma$ to quadratic order in the $\Phi$ field, and we can determine the tree-level masses for the pions. We get
\begin{equation}\label{eq:tree-level_masses1}
	m_{xy,t}^2 = 
	\mu (m_x + m_y) + a^2 \Delta(\xi_t)
\end{equation}
with $t$ the taste index, and $x$ and $y$ are any of the quarks. For $a=0$ we get the standard relationship between the pion mass squared and the quark masses. The additional term depends on the taste, $t$, and we have
\begin{eqnarray}\label{eq:deltas}
	\Delta (\xi_5) & \equiv & \Delta_P  = 0\nonumber \\*
	\Delta (\xi_{\mu5}) & \equiv & 
	\Delta_A = \frac{16}{f^2}\left( 
	C_1 + 3C_3 + C_4 + 3C_6 \right) \nonumber \\*
	\Delta (\xi_{\mu\nu})  & \equiv &\Delta_T =
	\frac{16}{f^2}\left(2C_3 + 2C_4 + 4C_6\right) \nonumber \\*
	\Delta (\xi_{\mu}) & \equiv & 
	\Delta_V = \frac{16}{f^2}\left( 
	C_1 + C_3 + 3C_4 + 3C_6 \right) \nonumber \\*
	\Delta (\xi_I)  & \equiv & \Delta_I =
	\frac{16}{f^2}\left( 
	4C_3 + 4C_4 \right).
\end{eqnarray}
Note that, to $\cO(a^2)$, there is a remnant $SO(4)$ symmetry that keeps certain tastes degenerate, and so we use labels that refer to the representations under $SO(4)$ rotations. This was first seen for the $N=1$ case in Ref.~\refcite{Lee:1999zx} and carries over to the general $N$ case.\cite{Aubin:2003mg} Also, only coefficients from $\cU$ contribute to the tree-level masses. $\cU^\prime$ does contribute here as well, but by giving additional two-point vertices much like the anomaly does. It only affects the flavor-neutral (diagonal) pions which have vector (V) or axial-vector (A) taste. Expanded to quadratic order in the fields, we have
\begin{equation}
	a^2\cU^\prime = \frac{a^2 \delta'_V}{2} (
	U_{\mu}+D_{\mu}+S_{\mu}+\cdots)^2 + V\to A
\end{equation}
where
\begin{equation}\label{eq:mix_vertex_V}
	\delta'_V \equiv \frac{16}{f^2} (C_{2V} - C_{5V})\ ,
\end{equation}
and similarly with $A$. These are of the same form as the anomaly term, and so we treat them in the same manner. The $\cU^\prime$ terms for the $V$ and $A$ tastes as well as the anomaly term for the singlet tastes all act in such a way to add off-diagonal terms to the mass matrix. By diagonalizing this matrix, we can write everything in terms of physical fields, $\pi^0_t,\eta_t,\eta'_t$.\footnote{We refer to these as physical because for the singlet case (relevant in the continuum) this amounts to changing from the flavor basis: $U,D,$ and $S$, to the physical basis: $\pi^0, \eta,$ and $\eta'$. Of course, the axial and vector taste fields are not strictly ``physical,'' but we will use the same terminology since they all result from diagonalizing the flavor-neutral mass matrix.} Of course, in the end, we send $m_0\to\infty$ since the $\eta'_I$ is heavy on the chiral scale, but we cannot do this for the $\eta'_A$ and $\eta'_V$. 

The procedure for diagonalizing the mass matrix, or equivalently calculating the full propagator, for the flavor-neutral fields is described quite well in Ref.~\refcite{Sharpe:2001fh}, and we just state the results here. For comparison, we first write the propagators for the flavor-charged fields with taste $t$, or flavor-neutrals with taste $T$ or $P$:
\begin{equation}
	G_{xy,t}(q^2) = \frac{1}
	{q^2 + m_{xy,t}^2}\ .
\end{equation}
For the flavor-neutrals, the propagator can be a little more complicated. For the taste-vector, taste-axial, or taste-singlet we have
\begin{equation}
	G_{xy,t}(q^2) = \frac{\delta_{xy}}
	{q^2 + m_{xx,t}^2}
	+\cD_{xy}^{t}(q^2)\ ,
\end{equation}
with for the axial and vector tastes:
\begin{eqnarray}\label{eq:discProp}
	\cD^{V(A)}_{xy}(q^2)
	&=&
	\frac{-a^2\delta'_{V(A)}}{(q^2 + m_{xx,V(A)}^2)
	(q^2 + m_{yy,V(A)}^2)}\nonumber\\*
	&&{}\times
	\frac{(q^2 + m_{uu,V(A)}^2)(q^2 + m_{dd,V(A)}^2)
	(q^2 + m_{ss,V(A)}^2)}
	{(q^2 + m_{\pi^0,V(A)}^2)(q^2 + m_{\eta,V(A)}^2)
	(q^2 + m_{\eta',V(A)}^2)}\ .
\end{eqnarray}
We have written \eq{discProp} explicitly for $N=3$, but one can see the pattern. There are $N$ factors of $(q^2+m^2)$ in both the numerator and denominator. Those in the numerator correspond to masses of the pions in the flavor basis and those in the denominator are those in the physical basis. The result is the same for the taste singlets, but we can take $m_0\to\infty$ in those terms to get
\begin{eqnarray}\label{eq:discPropSinglet}
	\cD^{I}_{xy}(q^2)
	&=&
	-\frac{1}{3}
	\frac{(q^2 + m_{uu,I}^2)(q^2 + m_{dd,I}^2)
	(q^2 + m_{ss,I}^2)}
	{(q^2 + m_{xx,I}^2)(q^2 + m_{yy,I}^2)
	(q^2 + m_{\pi^0,I}^2)(q^2 + m_{\eta,I}^2)}\ .
\end{eqnarray}

One final issue we have avoided thus far is the fact that we have an incorrect number of tastes per flavor in our theory. Our theory has $4N$ species of quarks, or four tastes for the $N$ flavors. The simplest procedure, for one-loop calculations, is to use a quark-flow analysis\cite{Sharpe:1992ft,Aubin:2003mg} to determine which quarks are ``sea'' quarks and thus correspond to quark loops. These are the loops which are removed by taking the fourth root in lattice simulations, and we must correct for this in the chiral theory. Since each quark loop can have four tastes, we merely multiply these terms by a factor of $1/4$ to reduce the theory to the correct number of tastes per flavor. 

The multiplication factor arises in two places. Terms which are connected (only include the $1/(q^2 + m^2)$ parts of the propagators) and those which are disconnected, coming from the $\cD$ portion of the flavor-neutral propagators. For the latter case, the correction is rather straightforward. As explained in detail in Ref.~\refcite{Aubin:2003mg}, we change the values of the parameters $\delta'_V$, $\delta'_A$, and $m^2_0$ that modify the masses of the flavor-neutral physical-basis pions by a factor of $1/4$. Explicit factors of these parameters remain unchanged, and this arises from the fact that the implicit factors have a one-to-one correspondence with the internal quark loops. The former case, which corresponds to the connected terms, is also rather straightforward. Identifying the quark loops leads to the conclusion that they must each be multiplied by a factor of $1/4$. This factor will be directly seen in our results unlike the correction to the disconnected pieces. More detail on quark flow analysis for the specific calculations to follow can be found in Ref.~\refcite{Aubin:2003mg,Aubin:2003uc,Aubin:2005aq}. 

At this point, we can calculate directly from the Lagrangian, \eq{ChiralAction}, quantities such as one-loop masses and decay constants, or pion scattering amplitudes. Of these, we will discuss the masses and decay constants. We can also calculate weak matrix elements, such as $B_K$, although this calculation is much more involved, so we will refer the reader to Ref.~\refcite{VandeWater:2005uq} for more details.

\subsection{Pion masses}

\begin{figure}
\begin{center}
 \includegraphics[width=5in]{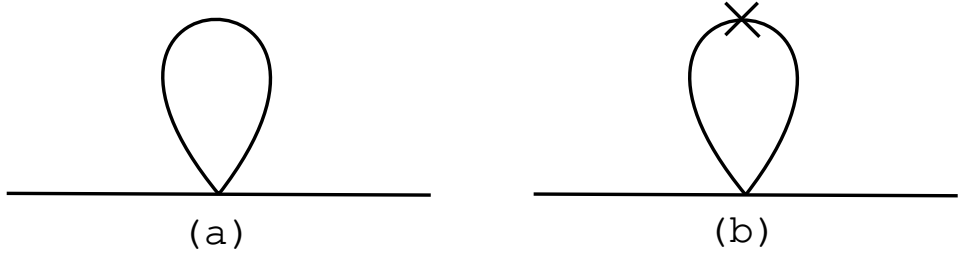} 
\end{center}
 \caption{The two one-loop 
 diagrams contributing to the self-energy of the pion, and thus to 
 its mass. (a) includes the connected diagrams and (b) includes the 
 disconnected diagrams.}
	\label{fig:Mass}
\end{figure}

To calculate the pion mass to one loop in \schpt, we need to expand out the Lagrangian to quartic order in $\Phi$ and use the four-point vertices that arise to calculate the diagrams shown in Fig.~\ref{fig:Mass}. These are the contributions to the self-energy of the pion field, and the first diagram includes those which are connected, meaning they come from the $1/(q^2+m^2)$ parts of propagators (both flavor-charged and flavor-neutral). The second diagram with the cross in the propagator includes those diagrams with only the flavor-neutral axial, singlet, and vector taste disconnected propagator, \cD.

We will also treat our calculation as a partially-quenched theory, where some quarks are treated as valence quarks while others are sea quarks. The valence quarks will not be allowed to propagate in loops. By using the same quark flow analysis to correct for the rooting trick, we can take these diagrams out by hand and not include them in our result. This is done in detail in Ref.~\refcite{Aubin:2003mg} for the pion masses, so we will not discuss this further here. The more formally correct method of partial quenching is to upgrade our symmetry group from $SU(4N)$ to $SU(4N_\textrm{sea}+4N_\textrm{val}|4N_\textrm{val})$,\cite{Bernard:1992mk,Bernard:1993sv} a graded group which can explicitly remove valence quark loop diagrams. For the cases discussed throughout this review, the quark loop method is simpler to employ, and leads to identical results.

To show only results that are specifically relevant to lattice simulations, we will pick $N=5$, where there are two valence quarks with masses $m_x$ and $m_y$, and three sea quarks with masses $m_l,m_l,$ and $m_s$. The up and down sea quarks are set to be degenerate, and the strange quark mass will be allowed to have a different mass. Picking out the ``real'' pion amounts to setting $m_x=m_y=m_l$ while the ``real'' kaon requires us to set $m_x=m_l$ and $m_y=m_s$. Also, we will only calculate the one-loop Goldstone pion, or the pseudoscalar-taste pion, as this is the pion whose mass vanishes in the chiral limit at finite lattice spacing (recall for this case, $\Delta_P=0$). 

The details of the calculation can be found in Ref.~\refcite{Aubin:2003mg}, and the result, including the tree-level analytic terms discussed above is
\begin{eqnarray}\label{eq:pionmass}
	\frac{\left(m^\textrm{1-loop}_{xy,P}\right)^2}{\mu(m_x+m_y)}
	& = &
	1 + \frac{1}{16\pi^2 f^2}\Biggl[
	\left(
	-2a^2\delta'_V\sum_{j} R^{[4,2]}_{j,V}\left(\cM^V_1;\cM^V_2\right)
	\ell(m^2_{j,V})\right)
	+\bigl( V\to A\bigr)
	\nonumber\\
	&&{}+
	\frac{2}{3}\sum_{j} R^{[3,2]}_{j,I}\left(\cM^I_3;\cM^I_2\right)
	\ell(m^2_{j,I})
	+\frac{16\mu}{f}(2L_8-L_5)(m_x+m_y)
	\nonumber\\
	&&{}+\frac{32\mu}{f}(2L_6-L_4)(2m_l+m_s)
	+a^2 C
	\Biggr]\ ,
\end{eqnarray}
with $L_i$ the Gasser-Leutwyler low-energy constants,\cite{Gasser:1985gg} $C$ is an unknown constant which comes from terms in $\cL_4$ that involve the lattice spacing,\cite{Sharpe:2004is} and 
\begin{equation}
	\ell(m^2) = m^2 \ln \left(\frac{m^2}{\Lambda^2}\right), \quad
	R_j^{[n,k]}
	\left(\left\{m\right\}\!;\!\left\{\mu\right\}\right)
    \equiv  
   \frac{\displaystyle\prod_{a=1}^k (\mu^2_a- m^2_j)}
    {\displaystyle\prod_{i=1\atop i\ne j}^n (m^2_i - m^2_j)}\ ,
\end{equation}
where $\Lambda$ is the chiral cutoff scale. The sets of masses given in the residues $R$ are those appearing in the numerators and denominators of the disconnected parts of the flavor-neutral propagators. More explicitly:
\begin{eqnarray*}
	&\cM^t_1 =  \{m_{xx,t},m_{yy,t},m_{\eta,t},m_{\eta',t} \} \ ,\
	\cM^t_2 = \{ m_{uu,t},m_{ss,t} \}\ ,&\\
	&\cM^I_3  =  \{m_{xx,I},m_{yy,I},m_{\eta,I} \}&\ .
\end{eqnarray*}
In \eq{pionmass}, the sums over $j$ run over the masses in the first argument of the residue functions $R$.

We will discuss the results from using this expression to fit lattice data after we calculate the decay constant.

\subsection{Pion decay constants}

\begin{figure}
\begin{center}
 \includegraphics[width=4in]{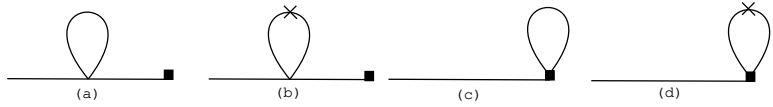} 
\end{center}
 \caption{The four one-loop 
 diagrams contributing to the decay constant of the pion, using the 
 same notation as Fig.~\ref{fig:Mass}. The box corresponds to an 
 insertion of the axial current. 
 (a) and (b) are the wavefunction renormalization diagrams (connected 
 and disconnected, respectively) and (c) and (d) are the vertex 
 corrections}
	\label{fig:Decay}
\end{figure}

The diagrams needed to calculate the pion decay constant, again in the partially-quenched case, are shown in Fig.~\ref{fig:Decay}, where the notation is the same as before. The addition here is the solid box, which represents an insertion of the axial current corresponding to the pseudoscalar taste pion, given at the chiral level by
\begin{eqnarray}\label{eq:current}
     j_{\mu5}^{xy,P} 
     &= &\frac{-i f^2}{8} 
     \Tr\left[\xi^{(N)}_5 \cP^{xy} \left(
     \partial_{\mu}\Sigma\Sigma^{\dagger} +
     \Sigma^{\dagger}\partial_{\mu}\Sigma
     \right)\right]\ ,
\end{eqnarray}
with $\cP^{xy}$ a projector which extracts the $4\times4$ block matrix which corresponds to the charged pion.\cite{Aubin:2003uc} We can extract the decay constant from the matrix element
\begin{equation}\label{eq:matrix_element}
        \left\langle 0 \left| j_{\mu5}^{xy,P}
        \right| P_5(p) \right\rangle =-i
        f_{xy,5} \; p_{\mu} \ ,
\end{equation}
with $P_5$ the pion state with quark content $xy$. With this normalization, $f_{\pi} \approx 131\ \MeV$. Note the correction coming from the wavefunction renormalization (and thus from the self-energy graphs above), are proportional to the vertex corrections.\cite{Aubin:2003uc}

The result for the decay constant is much more complicated than that of the mass:
\begin{eqnarray}\label{eq:final_21_result}
	\frac{f^\textrm{1-loop}_{xy,5}}{f}
	&& =  1 
	+ \frac{1}{16\pi^2 f^2}
	\Biggl[-\frac{1}{32}\sum_{q,t} 
	\left[
	\ell\left(m^2_{qx,t}\right)
	+\ell\left(m^2_{qy,t}\right)
	\right]\nonumber \\* &&
    - \frac{1}{6}\Biggl(
    \partial_{xx}^I
   	\sum_{j} 
	R^{[2,2]}_{j,I}\left(\cM^I_4;\cM^I_2\right)
	\})
    \ell(m^2_{j,I})
    \nonumber \\* &&
    +\partial_{yy}^{I}
    \sum_{j} 
    R^{[2,2]}_{j,I}\left(\cM^I_5;\cM^I_2\right)
    \ell(m^2_{j,I})
    +2\sum_{j}R^{[3,2]}_{j,I}\left(\cM^I_3;\cM^I_2\right)
    \ell(m^2_{j,I})
    \Biggr)\nonumber \\* &&
    +\frac{1}{2}a^2 \delta'_V\Biggl(  
    \partial_{xx}^{V}
    \sum_{j}
    R^{[3,2]}_{j,V}\left(\cM^V_6;\cM^V_2\right)
    \ell(m^2_{j,V})
    \nonumber \\* &&
    +\partial_{yy}^{V}
	\sum_{j}
	R^{[3,2]}_{j,V}\left(\cM^V_7;\cM^V_2\right)
	\ell(m^2_{j,V})
    -2\sum_{j}R^{[4,2]}_{j,V}\left(\cM^I_1;\cM^I_2\right)
    \ell(m^2_{j,V})
    \Biggr)      
    \nonumber \\* &&
    + \Bigl( V \to A \Bigr) \Biggr]
    + \frac{16\mu}{f^2}\left( 2m_{\ell} + m_s\right)L_4
    + \frac{8\mu}{f^2}\left( m_x + m_y \right)L_5
    +a^2 F \ ,
\end{eqnarray}
where we have used the shorthand notation
$\partial_{xx}^I \equiv \partial/\partial m_{xx,I}^2$,
and similarly for the other tastes and flavors. We have the additional mass sets
\begin{eqnarray*}
	\cM^t_4  =  \{m_{xx,t},m_{\eta,t} \}\ ,&\ \ &
	\cM^t_5  =  \{m_{yy,t},m_{\eta,t} \}\ , \\
	\cM^t_6  =  \{m_{xx,t},m_{\eta,t},m_{\eta',t} \}\ ,&\ \ &
	\cM^t_7  =  \{m_{yy,t},m_{\eta,t},m_{\eta',t} \}\ .
\end{eqnarray*}
Again, the sums over $j$ run over the mesons in the sets which are the first argument of the corresponding $R$'s. The sum over $t$ is over the 16 tastes and the sum over $q$ is over the sea quarks $u,d,$ and $s$. 

\subsection{Results from pion fits}

The MILC collaboration has fit to \eqs{pionmass}{final_21_result} with multiple lattice spacings ($a\approx 0.125\ \textrm{fm},0.09\ \textrm{fm},$ and $ 0.06\ \textrm{fm}$). As one can see from Refs.~\refcite{MILCSugar,Bernard:2006zp}, the fits are quite good, with a confidence level of 0.99. There are two subsets of fits, described in Ref.~\refcite{MILCSugar}, giving either 122 or 978 data points, and so the fits are constrained considerably. The most current results (still marked as preliminary), after taking the physical quark mass limit and the continuum limit are, for the decay constants
\begin{eqnarray*}
	&f_\pi = 128.6\pm0.4\pm3.0\ \MeV\ , \
	f_K = 155.3\pm0.4\pm3.1\ \MeV \ ,&\\
	&f_K/f_\pi  =  1.208(2){\textstyle \binom{+7}{-14}}\ ,&
\end{eqnarray*}
where the first errors are statistical and the second systematic, and for the quark masses (adding all the errors in quadrature)
\begin{eqnarray*}
	& m_s^{\overline{\textrm{MS}}} = 90(6)\ \MeV\  ,\ 
	m_l^{\overline{\textrm{MS}}}  =  3.3(2)\ \MeV\  &, \\
	&m_s/m_l = 27.2(4) \ .&
\end{eqnarray*}
One can also extract the Gasser-Leutwyler coefficients, the up and down quark masses, and certain CKM matrix elements, but we refer the reader to Ref.~\refcite{MILCSugar,Bernard:2006zp} for those quantities. 

The key point from these results is that with \eqs{pionmass}{final_21_result}, one can obtain rather precise results for certain quantities of interest with staggered fermions. Without these expressions, one would have no information about what the true chiral behavior of the pion mass and decay constant is near the continuum limit. In fact, for comparison, the confidence level of fits when using standard chiral expressions\cite{Aubin:2004fs} [\ie the continuum limit of \eqs{pionmass}{final_21_result}] is $\sim10^{-250}$. Using these expressions is crucial for a result with small, credible, errors.

\section{Heavy-lights}\label{sec:heavylights}

Having had success describing the simplest of quantities, we now turn our attention to more difficult quantities which are equally important for phenomenology. That is, including heavy-light mesons in our \schpt. With this, we can determine the chiral behavior of heavy-light decay constants, semileptonic form factors, as well as more complicated quantities such has $B\to D\ell\nu$ or $B\to D^*\ell\nu$ form factors.\cite{Aubin:2004xd,Aubin:2005aq,Laiho:2005ue}

The procedure for including heavy-light mesons is rather straightforward. We will sketch a few steps and refer the reader to Ref.~\refcite{Burdman:1992gh,Boyd:1994pa,Aubin:2005aq} for more details, both on the continuum and staggered version of this procedure. After doing this, we will calculate the heavy-light decay constant for a $B$ meson.

The first step is incorporate heavy quark effective theory (HQET) into \schpt. We take the masses of the heavy quarks to be large (compared to $\Lambda_{QCD}$) and perform a $1/m_Q$ expansion. For this review, we will only keep leading order terms in this expansion [$\cO(1)$ terms]. A heavy-light field, $H^{Q}_a$, has two indices, one we'll denote as a superscript $Q$ to refer to the heavy quark flavor and the other will be $a$ to denote the light quark flavor/taste. We also assume that the heavy quark is discretized in such a way that the doublers have large enough masses so they don't affect the dynamics (see Ref.~\refcite{Aubin:2005aq} for a more detailed discussion of this restriction), and so we treat it as a continuum quark. 

For $m$ heavy quarks, we have a combined spin-flavor $SU(2m)$ symmetry in addition to the chiral $SU(4N)_L\times SU(4N)_R$ symmetry. In matrix notation, we have
\begin{equation}
	H \to S H \mathbb{U}^\dag,
	\quad \overline{H} \to \mathbb{U} 
	\overline{H} S^\dag\ ,
\end{equation}
where $S\in SU(2m)$, and $\mathbb{U}\in SU(4N)$. We pick a special transformation for the right-hand side, so as to keep $H$'s transformation under parity simple. This matrix $\mathbb{U}$ is spacetime dependent and is defined with $\sigma^2 = \Sigma$, where $\sigma$ transforms under the chiral symmetry group as
\begin{equation}
	\sigma\to L\sigma \mathbb{U}^\dag = \mathbb{U}\sigma R\ .
\end{equation}

When we enumerate all of the possible operators that is part of the heavy-light \schpt\ action, we find there are over 200 terms in addition to the continuum action! All of these additional terms, however, contribute in the same way only to analytic terms when working to one-loop order in any given quantity. As these terms are all proportional to $a^2$, however, they are unphysical quantities; thus it is not necessary to determine precise values for them all. For the following calculation, we will just write down the relevant action required for the one-loop terms which involve the heavy-lights:
\begin{equation}\label{eq:hlaction}
	\cL_{hl} = 
	-i \Tr\left(\overline{H}H v\cdot \leftvec D\right)
	+ g_\pi\Tr\left(\overline{H}H\gamma^\mu\gamma_5\mathbb{A}_\mu
	\right)\ ,
\end{equation}
where the trace is a combined trace over the heavy quark indices, the light quark flavor-taste indices and the Dirac indices, and 
\begin{eqnarray}
	\left(H \leftvec D_\mu\right)_a & = &
	\partial_\mu H_a + i H_b \mathbb{V}^{ab}_\mu \ , \\
	\mathbb{V}_\mu & \equiv &\frac{i}{2}\left[
	\sigma^\dag\partial_\mu\sigma + \sigma\partial_\mu\sigma^\dag
	\right]\ , \\
	\mathbb{A}_\mu & \equiv &\frac{i}{2}\left[
	\sigma^\dag\partial_\mu\sigma - \sigma\partial_\mu\sigma^\dag
	\right]\ .
\end{eqnarray}
One can verify that \eq{hlaction} is invariant under the full HQET-\schpt\ symmetry group.\footnote{Note that \eq{hlaction} is in Minkowski space, because it is more consistent with other (continuum) work with heavy-lights. By changing the signs of the mass term and $\cV$ in \eq{ChiralAction}, one can obtain the \schpt\ action in Minkowski space as well.}
There are terms which arise due to the finite light quark masses and the lattice spacings, but these will only contribute to analytic terms, and so we will not include them here. 

\begin{figure}
\begin{center}
 \includegraphics[width=5in]{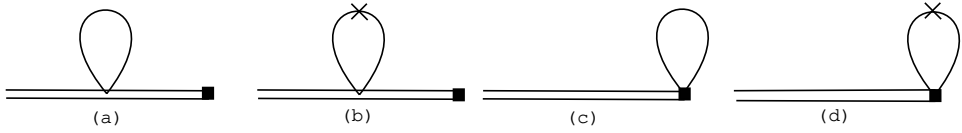} 
\end{center}
 \caption{The four one-loop 
 diagrams contributing to the $B$ decay constant. 
 (a) and (b) are the wavefunction renormalization diagrams (connected 
 and disconnected, respectively) and (c) and (d) are the vertex 
 corrections.}
	\label{fig:fB}
\end{figure}

To calculate $f_B$, we need the left-handed current. To leading order, this takes the form
\begin{equation}\label{eq:hlcurrent}
	j^{\mu, c} = \frac{\kappa}{2}
	\tr_D\left(\gamma^\mu(1-\gamma_5)H\right)
	\sigma^\dag\lambda^{(c)}\ ,
\end{equation}
where $\lambda^{(c)}$ is a constant vector which fixes the taste-flavor: $\left(\lambda^{(c)}\right)_a = \delta_{ac}$ and $\tr_D$ is a trace over only Dirac indices. We have not included an index on the heavy-light field for the heavy quark content, although below we will refer to a $B$ meson, a heavy-light made from a bottom quark and a light quark, to stress the fact that we are working to leading order in the heavy quark mass. The decay constant is defined through the matrix element
\begin{equation}\label{eq:hldecay}
	\left\langle 0 | j^{\mu, c} | B_a(v) \right\rangle
	=
	if_{B_a} m_{B_a} v^\mu \delta_{ac}\ ,
\end{equation}
where the normalization of the $B$ state is given in Ref.~\refcite{Aubin:2005aq}. At leading order we have $f^\textrm{LO}_{B_a} = \kappa/\sqrt{m_{B_a}}$. The non-vanishing diagrams are shown in Fig.~\ref{fig:fB}, where the notation is the same as in the pion case, except we now have heavy-light mesons, denoted by the double lines. We need to account for the factors of 1/4 to correct for the fourth-root trick, and it turns out through a careful quark-flow analysis\cite{Aubin:2005aq} to arise in the same way as for pion quantities. The results below have already taken this into account.

We have for the partially quenched decay constant for 2+1 flavors
\begin{eqnarray}\label{eq:2p1_pq_fB}
  \frac{f_{B_x}}{f_{B_x}^{\rm LO}} 
  & = & 1 + \frac{1}{16\pi^2f^2}
  \frac{1+3g_\pi^2}{2}
  \Biggl\{-\frac{1}{16}\sum_{q,t} \ell(m_{xq,t}^2)
  - \frac{1}{3}
	\sum_j \partial_{xx}^{I}\left[ 
      R^{[2,2]}_{j}(
    \cM_4^I;  \cM_{2}^I) \ell(m_{j}^2) \right]
    \nonumber \\*&&{} 
     -   \biggl( a^2\delta'_V \sum_{j}
     \partial_{xx}^{V}\left[ 
       R^{[3,2]}_{j}( \cM^V_6; \cM_{2}^V)
    \ell(m_{j}^2)\right]
        + [V\to A]\biggr)  
   \Biggr\} \nonumber \\*&&{}+
        c_s (2m_l + m_s) + c_v m_x + c_a a^2 \ ,
\end{eqnarray}
where the sums over $t,q,$ and $j$ are the same as in the pion case, and the mass sets are defined above. The analytic terms are combinations of numerous unknown parameters from the Lagrangian as well as higher-order terms in the current. Finally, due to a shift symmetry in the HQET-\schpt\ Lagrangian, this result is independent of the taste of the light quark in the $B$ meson.\cite{Aubin:2004xd}

Although we have this calculation in terms of the $B$ meson, in principle it is also valid to leading order for the $D$ meson, although some errors may be larger due to the smaller mass of the charm quark. Nevertheless, the $D$ and $D_s$ decay constants have been measured by the Fermilab Lattice group:\cite{Aubin:2005ar}
\begin{equation}\label{eg:fDresults}
	f_{D^+} = 201(3)(17)\ \MeV\ ,\ \
	f_{D_s} = 249(3)(16)\ \MeV\ ,
\end{equation}
which agrees with the experimental measurement that was posted shortly afterwards.\cite{Artuso:2005ym} Again, from these results, one can extract CKM matrix elements.\cite{Aubin:2005ar}

Using this HQET-\schpt\ theory, we can also calculate the semileptonic form factors for $B\to \pi\ell\nu$ decay, $B\to D^*\ell\nu$, or many other quantities one might be interested in calculating on the lattice. We do not discuss these results in detail, but instead refer the reader to Ref.~\refcite{Aubin:2004xd,Aubin:2005aq,Laiho:2005ue} for details.

\section{Conclusions}\label{sec:conc}

We have discussed many of the interesting applications of \schpt\ to extract physical results from lattice data. There are many other quantities we have not discussed, but would like to touch upon here. Van de Water \& Sharpe have analyzed the neutral kaon mixing parameter, $B_K$, using \schpt.\cite{VandeWater:2005uq} This is rather extensive and requires a large number of additional parameters to control all of the errors that arise in a staggered calculation of $B_K$ (operator mixing, perturbative errors, \etc). Perhaps this is an indication of the practical limitations of staggered quarks, as when the quantities become more complicated (as is the case with weak matrix elements), the difficulty increases dramatically. An interesting solution for $B_K$ and other weak matrix elements could be a ``mixed-action'' approach, using different valence quarks on top of staggered sea quarks.\cite{Bar:2002nr,Bar:2005tu,Aubin:2006hg}

A slightly different application is to that of a calculation of the muon $g\!-\!2$.\cite{Blum:2002ii,Blum:2003se,Blum:2004cq,Aubin:2006xv} A calculation of the muon $g\!-\!2$ on the lattice is actually a calculation of the photon vacuum polarization, and using \schpt\ combined with electromagnetic interactions for this can aid in extracting the low-energy behavior of this function. For this, an extension of \schpt\ to include light resonances (mainly the vectors) is needed, as they dominate the low-energy behavior. Although this has been used successfully to describe the lattice results, the errors are not yet competitive enough with other techniques to calculate the muon $g\!-\!2$ theoretically (see Ref.~\refcite{Hagiwara:2006jt} for example).

In conclusion, at least for the simpler quantities, precise staggered simulations combined with \schpt\ expressions allow for accurate determinations of physical quantities that can be compared against experiment reliably. This is an indication that lattice QCD can be used in the forthcoming years as a spectacular testing ground for more precise determinations of the effects of the strong interactions. Whether or not more complicated quantities can be adequately determined using staggered quarks is an open question, but for the present, the use of \schpt\ has made possible these precise extrapolations.

\section*{Acknowledgments}

This work was supported by the U.S.\ Department of Energy.

\end{document}